
\magnification=1200
\baselineskip=18pt

\def\[{\c c}
\def\\{\'\i }

\font\tituloa=cmb10 scaled\magstep1
\font\titulob=cmb10 scaled\magstep2

\bigskip
\centerline{\titulob Geometric quantization of topological gauge
theories}
\vskip 2.0 cm
\centerline{J. Barcelos-Neto$\,^\ast$ and S.M. de Souza}
\bigskip
\centerline{\it Instituto de F\\sica}
\centerline{\it Universidade Federal do Rio de Janeiro}
\centerline{\it RJ 21945-970 - Caixa Postal 68528 - Brasil}
\vskip 2 cm
\centerline{\tituloa Abstract}
\bigskip
We study the symplectic quantization of Abelian gauge theories in
$2+1$ space-time dimensions with the introduction of a topological
Chern-Simons term.

\vfill
\bigskip
\hbox to 3.5 cm {\hrulefill}\par
\item{($\ast$)} Electronic mail: ift03001 @ ufrj
\eject

{\tituloa 1. Introduction}
\bigskip
Gauge theories developed in (2+1) spacetime dimensions have
interesting and intriguing properties. This occurs due to the
introduction of a topological term in the Lagrangian in order that
the theory can be completely formulated. These terms are called
Chern-Simons~(CS) in the literature~[1].

\medskip
Among the above-mentioned features we refer to the generation of a
(topological) mass for the gauge field~[1,2] and the possibility of
appearing exotic statistics when there is a coupling with matter
fields~[3]. More recently, the interest in CS theories has grown up by
virtue of a Witten work~[4] where it was shown the connection between
three-dimensional topological field theories and conformal field
theories in two dimensions.

\medskip
On the other hand, one can say that quantization of CS theories is a
fascinating subject by its own right. This occurs due to its peculiar
structure of constraints. The canonical quantization in the
non-covariant Weyl (temporal) and Coulomb gauges was first achieved
by Deser et al.~[5]. A manifest covariant canonical quantization was
carried out more recently~[6] and also the quantization by means of
operator formalism~[7]. The canonical quantization based on the Dirac
Hamiltonian procedure~[8] was developed by Lin and Ni~[9], using the
temporal gauge, and Mart\\nez-Fern\'andes and Wotzasek~[10], using
the Coulomb gauge.

\medskip
The purpose of the present work is to use the symplectic
formalism~[11,12] to quantize this interesting constrained
system~[13,14]. We give the following organization to our paper: In
Sec. 2 we deal with the pure CS theory. In Sec. 3 we consider Maxwell
plus CS. Sec. 4 contains some concluding remarks.

\vskip 1cm
{\tituloa 2. Pure Abelian CS theory}
\bigskip
The pure CS Lagrangian density is

$${\cal L}={\kappa\over4\pi}\,
\epsilon^{\mu\nu\rho}\,\partial_\mu A_\nu A_\rho\eqno (2.1)$$

\bigskip
\noindent where $\kappa$ is a dimensionless coupling constant. We
adopt the following conventions $\eta_{\mu\nu}={\rm diag.}(1,-1,-1)$
and $\epsilon_{012}=\epsilon^{012}=1$.

\bigskip
Developing the CS Lagrangian we may write

$${\cal L}^{(0)}={\kappa\over4\pi}\,
\epsilon^{ij}A_j\dot A_i-V^{(0)}\eqno(2.2)$$

\bigskip
\noindent where the superscript $^{(0)}$ means an initial Lagrangian,
i.e., without the introduction of constraint terms. The quantity
$\epsilon^{ij}$ is the projection of $\epsilon^{\mu\nu\rho}$ in the
subspace spanned by space indices and

$$V^{(0)}= -{\kappa\over2\pi}\,
\epsilon^{ij}\,\partial_iA_jA_0\eqno(2.3)$$

\bigskip
\noindent From expression (2.2) we identify the symplectic
coefficients~[11]

$$\eqalignno{a^{(0)i}(\vec x,t)&={\kappa\over4\pi}\,
\epsilon^{ij}A_j(\vec x,t)\cr
a^{(0)0}(\vec x,t)&=0&(2.4)\cr}$$

\bigskip
\noindent This permit us to calculate the matrix elements

$$\eqalignno{f^{(0)ij}(\vec x,\vec y)&=
{\delta a^{(0)j}(\vec y)\over\delta A_i(\vec x)}
-{\delta a^{(0)i}(\vec x)\over\delta A_j(\vec y)}\cr
&=-{\kappa\over2\pi}\,
\epsilon^{ij}\,\delta(\vec x-\vec y)\cr
f^{(0)0j}(\vec x,\vec y)&=0
=f^{(0)00}(\vec x,\vec y)&(2.5)\cr}$$

\bigskip
\noindent Now and throughout it will be understood that all
quantities are taken at the same time. The matrix $f^{(0)}$ reads

$$f^{(0)}=\pmatrix{\strut0&0\cr
0&-{\strut\displaystyle\kappa\epsilon^{ij}
\over\strut\displaystyle2\pi}\cr}\,
\delta(\vec x-\vec y)\eqno(2.6)$$

\bigskip
\noindent which is obviously singular. This means that this system
has constraints. Let us consider that the general form of the
eigenvector with zero eigenvalue is~[13,14]

$$v^{(0)}=\pmatrix{v^{(0)}_0\cr
v^{(0)}_j\cr}\eqno(2.7)$$

\bigskip
\noindent We notice that this is actually true if

$$v^{(0)}_j=0\eqno(2.8)$$

\bigskip
\noindent Possible constraints are obtained from

$$\int d^2\vec x\,\Bigl(v^{(0)0}(\vec x)\,
{\delta\over\delta A^0(\vec x)}
+v^{(0)i}(\vec x)\,{\delta\over\delta A^i(\vec x)}\Bigr)\,
\int d^2\vec y\,V^{(0)}=0\eqno(2.9)$$

\bigskip
\noindent Considering the expression for $V^{(0)}$ given by (2.3) and
the condition given by eq. (2.8), expression (2.9) leads to

$$\epsilon^{ij}\int d^2\vec x\,v^{(0)0}(\vec x)\,
\partial_iA_j(\vec x)=0\eqno(2.10)$$

\bigskip
\noindent Since $v^{(0)0}$ is an arbitrary function of $\vec x$ we
conclude that

$$\epsilon^{ij}\,\partial_iA_j(x)=0\eqno(2.11)$$

\bigskip
The next step is to introduce this constraint into the kinetic part
of the Lagrangian [13, 14]. However, looking at (2.3) we notice that
it is already in the potential part. Then, what we have to do is just
to transpose it from the potential to the kinetic sector. This is
directly done by making

$$A_0=\dot\lambda\eqno(2.12)$$

\bigskip
\noindent After this replacement the Lagrangian turns to be

$${\cal L}^{(1)}={\kappa\over4\pi}\,\epsilon^{ij}A_j\dot A_i
+{\kappa\over2\pi}\,\epsilon^{ij}\partial_iA_j\dot\lambda
\eqno(2.13)$$

\bigskip
\noindent The potential $V^{(1)}$ is obviously zero. From the
Lagrangian above one identifies the coefficients

$$\eqalignno{a^{(1)i}(x)&={\kappa\over4\pi}\,
\epsilon^{ij}A_j(x)\cr
a^{(1)}\,_\lambda(x)&={\kappa\over2\pi}\,
\epsilon^{ij}\,\partial_iA_j(x)&(2.14)\cr}$$

\bigskip
\noindent This leads to the matrix $f^{(1)}$

$$\eqalignno{f^{(1)}&=\pmatrix{f^{(1)ij}&f^{(1)i}\,_\lambda\cr
f^{(1)}\,_\lambda\,^j&f^{(1)}\,_{\lambda\lambda}\cr}\cr
&={\kappa\over2\pi}\,\pmatrix{-\epsilon^{ij}
&\epsilon^{ik}\partial_k\cr
\epsilon^{jk}\partial_k&0\cr}\,
\delta(\vec x-\vec y)&(2.15)\cr}$$

\bigskip
\noindent Without explicit indication, partial derivatives will
always be understood to be acting on the variable $\vec x$.

\medskip
The matrix above is still singular
\footnote{(*)}{In order to confirm this fact one can show that there
actually exists a nontrivial eigenvector with zero eigenvalue.
Considering that a general form of this eigenvector is
$v^{(1)}=(v^{(1)}_j,v^{(1)}_\lambda)$, this will be a zero mode with
the condition $v^{(1)}_i-\partial_iv^{(1)}_\lambda=0$.}.
Further, there is no possibility to find out new constraints because
as we have seen $V^{(1)}=0$. Now is the point where the gauge
condition has to be chosen. Let us first consider the temporal gauge

$$A_0(x)=0\eqno(2.16)$$

\bigskip
\noindent which means from (2.12) that $\lambda=$ constant. We then
introduce this new constraint into
\vfill\eject

the kinetic part of the Lagrangian
by means of a Lagrange multiplier $\eta$. The result is

$${\cal L}^{(2)}={\kappa\over4\pi}\,
\epsilon^{ij}\,A_j\dot A_i
+{\kappa\over2\pi}\,\Bigl(\epsilon^{ij}\,
\partial_iA_j+\eta\Bigr)\,\dot\lambda\eqno(2.17)$$

\bigskip
\noindent Thus, the new coefficients are

$$\eqalignno{a^{(2)i}&={\kappa\over4\pi}\,\epsilon^{ij}\,A_j\cr
a^{(2)}\,_\lambda&={\kappa\over2\pi}\,\Bigl(\epsilon^{ij}\,
\partial_iA_j+\eta\Bigr)\cr
a^{(2)}\,_\eta&=0&(2.18)\cr}$$

\bigskip
\noindent which leads to the matrix
$$f^{(2)}={\kappa\over2\pi}\,
\pmatrix{-\epsilon^{ij}&\epsilon^{ik}\partial_k&0\cr
\epsilon^{jk}\partial_k&0&-1\cr
0&1&0\cr}\,\delta(\vec x-\vec y)\eqno(2.19)$$

\bigskip
\noindent where rows and columns follow the order $A^i$, $\lambda$
and $\eta$. The above matrix is not singular and, consequently, it
can be identified as the  symplectic tensor. Its inverse reads

$$f^{(2)^{-1}}={2\pi\over\kappa}\,
\pmatrix{\epsilon_{ij}&0&\partial_i\cr
0&0&1\cr\partial_j&-1&0\cr}\,
\delta(\vec x-\vec y)\eqno(2.20)$$

\bigskip
\noindent The Dirac brackets of the theory correspond to the elements
of this inverse matrix [11, 13, 14]. We thus have

$$\eqalignno{\bigl\{A_i(\vec x), A_j(\vec y)\bigr\}
&={2\pi\over\kappa}\,\epsilon_{ij}\,\delta(\vec x-\vec y)\cr
\bigl\{A_i(\vec x),\eta(\vec y)\bigr\}
&={2\pi\over\kappa}\,\partial_i\,\delta(\vec x-\vec y)\cr
\bigl\{\lambda(\vec x),\eta(\vec y)\bigr\}
&={2\pi\over\kappa}\,\delta(\vec x-\vec y)&(2.21)\cr}$$

\bigskip
\noindent The first bracket above was the same found Lin and Ni~[9].

\medskip
Let us next choose the Coulomb gauge

$$\partial^iA_i=0\eqno(2.22)$$

\bigskip
\noindent Taking the time derivative of this constraint and
introducing the result into the Lagrangian (2.13) by means of a
Lagrange multiplier, we get

$${\cal L}^{(2)}={\kappa\over2\pi}\,
\Bigl({1\over2}\epsilon^{ij}A_j+\partial^i\zeta\Bigr)\dot A_i
+{\kappa\over2\pi}\,\epsilon^{ij}\partial_iA_j\dot\lambda
\eqno(2.23)$$

\bigskip
\noindent Following the same previous procedure we find the matrix

$$f^{(2)}={\kappa\over2\pi}\,
\pmatrix{-\epsilon^{ij}&\epsilon^{ik}\partial_k&-\partial^i\cr
\epsilon^{jk}\partial_k&0&0\cr
-\partial^j&0&0\cr}\,\delta(\vec x-\vec y)\eqno(2.24)$$

\bigskip
\noindent whose inverse is

$$f^{(2)^{-1}}={2\pi\over\kappa}\,
\pmatrix{0&-{\strut\displaystyle\epsilon_{jk}\partial^k\over
\strut\displaystyle\nabla^2}
&{\strut\displaystyle\partial_j\over\strut\displaystyle\nabla^2}\cr
-{\strut\displaystyle\epsilon_{ik}\partial^k\over
\strut\displaystyle\nabla^2}&0
&{\strut\displaystyle1\over\strut\displaystyle\nabla^2}\cr
{\strut\displaystyle\partial_k\over\strut\displaystyle\nabla^2}
&-{\strut1\over\displaystyle\nabla^2}&0\cr}\,
\delta(\vec x-\vec y)\eqno(2.25)$$

\bigskip
\noindent where rows and columns follow the order $A^i$, $\lambda$
and $\zeta$. From (2.25) we identify the nonvanishing brackets

$$\eqalignno{\bigl\{A_i(\vec x),\lambda(\vec y)\bigr\}
&=-{2\pi\over\kappa}{\epsilon_{ik}\partial^k\over\nabla^2}\,
\delta(\vec x-\vec y)\cr
\bigl\{A_i(\vec x),\zeta(\vec y)\bigr\}
&={2\pi\over\kappa}{\partial_i\over\nabla^2}\,
\delta(\vec x-\vec y)\cr
\bigl\{\lambda(\vec x),\zeta(\vec y)\bigr\}
&={2\pi\over\kappa}{1\over\nabla^2}\,
\delta(\vec x-\vec y)&(2.26)\cr}$$

\bigskip
\noindent Now, the bracket $\{A_i(\vec x),A_j(\vec y)\}=0$ is zero.
This result is in agreement with the one found in~[10].

\medskip
To conclude this section we mention that the use of the axial gauge,
namely,

$$A_2\approx0\eqno(2.27)$$

\bigskip
\noindent gives

$$\eqalignno{\bigl\{A_i(\vec x), A_j(\vec y)\bigr\}
&=\bigl(\epsilon_{ij}-\epsilon_{ik}\delta_2^k
+\epsilon_{jk}\delta_2^k\bigr)\,
\delta(\vec x-\vec y)\cr
&=0&(2.28)\cr}$$

\bigskip
\noindent and the following ones  involving Lagrange multipliers

$$\eqalignno{\bigl\{A_i(\vec x),\lambda(\vec y)\bigr\}
&={2\pi\over\kappa}\,\epsilon_{i2}\partial_2^{-1}\,
\delta(\vec x-\vec y)\cr
\bigl\{A_i(\vec x),\zeta(\vec y)\bigr\}
&={4\pi\over\kappa}\,\partial_i\partial_2^{-1}\,
\delta(\vec x-\vec y)\cr
\bigl\{\lambda(\vec x),\zeta(\vec y)\bigr\}
&={4\pi\over\kappa}\,\partial_2^{-1}\,
\delta(\vec x-\vec y)&(2.29)\cr}$$

\vskip 1cm
{\tituloa 3. Maxwell plus CS}
\bigskip
The initial Langrangian in this case is

$${\cal L}^{(0)}=-{1\over4}\,F_{\mu\nu}F^{\mu\nu}
+{\kappa\over4\pi}\,\epsilon^{\mu\nu\rho}\,
\partial_\mu A_\nu A_\rho\eqno(3.1)$$

\bigskip
\noindent Using the momentum $\pi^\mu$ conjugate to $A_\mu$ as an
auxiliary field to linearize the Lagrangian, we get

$${\cal L}^{(0)}= \pi^i\dot A_i-V^{(0)}\eqno(3.2)$$

\bigskip
\noindent where

$$\eqalignno{V^{(0)}=-{1\over2}\pi^i\pi_i
&-{\kappa^2\over32\pi^2}\,A^iA_i
+\Bigl(\partial_iA_0+{\kappa\over4\pi}\,
\epsilon_{ij}A^j\Bigr)\,\pi^i\cr
&+{\kappa\over4\pi}\,\epsilon_{ij}
\partial^iA^0A^j
+{1\over2}\,\Bigl(\epsilon^{ij}\partial_iA_j\Bigr)^2
&(3.3)\cr}$$

\bigskip
\noindent This permit us to obtain the matrix

$$f^{(0)}=\pmatrix{0&0&0\cr
0&0&\delta^{ij}\cr
0&-\delta^{ij}&0\cr}\,
\delta(\vec x-\vec y)\eqno(3.4)$$

\bigskip
\noindent Here, rows and columns follow the order $A_0$, $A_i$ and
$\pi_i$. This matrix is singular and the corresponding zero mode gives
the constraint

$$\partial^i\,\Bigl(\pi_i
+{\kappa\over4\pi}\,\epsilon_{ij}A^j\Bigr)
\approx0\eqno(3.5)$$

\bigskip
\noindent Introducing it into the kinetic part of ${\cal L}^{(0)}$ by
means of a Lagrange multiplier we have

$${\cal L}^{(1)}=\Bigl(\pi_i+{\kappa\over4\pi}\,
\epsilon_{ij}\partial^j\lambda\Bigr)\,\dot A^i
+\partial_i\lambda\,\dot\pi^i-V^{(1)}\eqno(3.6)$$

\bigskip
\noindent where

$$V^{(1)}=-{1\over2}\pi^i\pi_i
+{\kappa^2\over32\pi^2}\,A^iA_i
-{\kappa\over4\pi}\,\epsilon_{ij}A^j\,\pi^i
-{1\over2}\,\Bigl(\epsilon^{ij}\partial_iA_j\Bigr)^2
\eqno(3.7)$$

\bigskip
\noindent $A_0$ has been absorbed in $\dot\lambda$. Now, the
corresponding matrix is

$$f^{(1)}=\pmatrix{0&\delta^{ij}
&{\strut\displaystyle\kappa\over\strut\displaystyle4\pi}\,
\epsilon^{ik}\partial_k\cr
-\delta^{ij}&0&\partial^i\cr
{\strut\displaystyle\kappa\over\strut\displaystyle4\pi}\,
\epsilon^{ik}\partial_k
&\partial^j&0\cr}\,\delta(\vec x-\vec y)\eqno(3.8)$$

\bigskip
\noindent which is also singular. However the zero modes do not
generate new constraints.

\medskip
As we have done in the previous section, let us fix the gauge.
Following the same procedure as before we have that for the temporal
gauge the result is

$$\eqalignno{\bigl\{A_i(\vec x), A_j(\vec y)\bigr\}&=0\cr
\bigl\{A_i(\vec x), \pi_j(\vec y)\bigr\}
&=-\delta_{ij}\,\delta(\vec x-\vec y)&(3.9)\cr}$$

\bigskip
\noindent Plus other brackets involving the Lagrange multipliers. The
result above is also in agreement with the work of ref.~[9].
Considering now the Coulomb gauge, the brackets just involving $A_i$ and
$\pi_j$ are

$$\eqalignno{\bigl\{A_i(\vec x), A_j(\vec y)\bigr\}&=0\cr
\bigl\{A_i(\vec x), \pi_j(\vec y)\bigr\}
&=-\Bigl(\delta_{ij}-{\partial_i\partial_j\over\nabla^2}
\Bigr)\,\delta(\vec x-\vec y)&(3.10)\cr}$$

\bigskip
\noindent that are also in agreement with the results found in~[10].
Finally, for the axial gauge, the result is

$$\eqalignno{\bigl\{A_i(\vec x), A_j(\vec y)\bigr\}&=0\cr
\bigl\{A_i(\vec x), \pi_j(\vec y)\bigr\}
&=\bigl(-\delta_{ij}+\delta_{2j}\partial_i\partial^{-1}_2
\bigr)\,\delta(\vec x-\vec y)&(3.11)\cr}$$

\vskip 1cm
{\tituloa 4. Quantization and propagators}
\bigskip
Since there are no problem with ordering operators, all the previous Dirac
brackets, obtained by means of the  symplectic formalism, can be
transformed to commutators by means of the usual rule
$\{Dirac\,brackets\}\,\rightarrow\,-i\hbar\,[commutators]$.
With these quantum relations one can calculate the propagators. We
list below the results (we just consider propagators among gauge
fields)

\bigskip\medskip
(i) {\bf CS and axial gauge}
\bigskip
$$G_{ij}(k)=0\eqno(4.1)$$

\bigskip\medskip
(ii) {\bf CS and temporal gauge}
\bigskip
$$G_{ij}(k)=i{2\pi\over\kappa}\,
{\epsilon_{ij}\over k_0}\eqno(4.2)$$

\bigskip\medskip
(iii) {\bf CS and Coulomb gauge}
\bigskip
$$G_{ij}(k)=0\eqno(4.3)$$

\bigskip\medskip
(iv) {\bf Maxwell plus CS and axial gauge}
\bigskip
$$G_{11}(k)={1\over\strut\displaystyle
k^2-{\kappa^2\over\strut\displaystyle4\pi^2}}\,
\Bigl(1+{k_1k_2\over k_2^2}\Bigr)\eqno(4.4)$$

(v) {\bf Maxwell plus CS and temporal gauge}
\bigskip
$$G_{ij}(k)={1\over k^2
-{\strut\displaystyle\kappa^2
\over\strut\displaystyle4\pi^2}}\,
\biggl(\delta_{ij}-{k_ik_j\over k^2_0}
-{i\kappa k_0\over2\pi k^2}\,
\Bigl(\epsilon_{ij}-{\epsilon_{jl}k_ik^l\over k_0^2}
+{\epsilon_{il}k_jk^l\over k_0^2}\Bigr)\biggr)\eqno(4.5)$$

\bigskip
\medskip
(vi) {\bf Maxwell plus CS and Coulomb gauge}
\bigskip

$$G_{ij}={1\over k^2
-{\strut\displaystyle\kappa^2
\over\strut\displaystyle4\pi^2}}\,
\biggl(\delta_{ij}-{k_ik_j\over\vec k^2}
-{i\kappa k^0\over2\pi\bigl(k^2
-{\strut\displaystyle\kappa^2
\over\strut\displaystyle4\pi^2}\bigr)}\,
\Bigl(\epsilon_{ij}-{\epsilon_{il}k^lk_j\over\vec k}
+{\epsilon_{jl}k^lk_i\over\vec k}\Bigr)\biggr)\eqno(4.6)$$

\vskip 1cm
{\tituloa 4. Conclusion}
\bigskip
We have studied the quantization of Abelian gauge theories in 2+1
space-time dimensions by using the symplectic formalism. These
theories, developed in odd dimensions, have the inclusion of a
topological term in order to be consistently defined. The presence
of this term gives us an interesting constrained system where the
symplectic method could be verified. The results we have
obtained are in agreement with those one previously obtained by means
of the standard Dirac procedure.

\medskip
In this paper we have just used Abelian gauge fields. Extensions to
nonabelian ones can be done in a straightforward way without any
great difficulty and the results do not present any special feature
that justify to be presented here.

\vskip 1cm
{\tituloa Acknowledgment}
\bigskip
We are in debt with C. Wotzasek for many useful comments. This
work was supported in part by Conselho Nacional de Desenvolvimento
Cient\\fico e Tecnol\'ogico - CNPq (Brazilian Research Agency).

\vfill\eject
{\tituloa References}
\bigskip
\item {1.} J.F. Schonfeld, Nucl. Phys. B185 (1981) 157; R. Jackiw and
S. Templeton, Phys. Rev. D23 (1981) 2291.
\item {2.} S. Deser, R. Jackiw and S. Templeton, Phys. Rev. Lett. 48
(1982) 975.
\item {3.} C.R. Hagen, Ann. Phys. 157 (1984) 342. For a general
review see R. Mac Kenzie and F. Wilczek, Int. J. Mod. Phys. A3 (1988)
2827 and P. De Sousa Gerbert, Int. J. Mod. Phys. A6 (1991) 173, and
references therein.
\item {4.} E. Witten, Commun. Math. Phys. 121 (1989) 351.
\item {5.} S. Deser, R. Jackiw and S. Templeton, Ann. Phys. 140
(1982) 372.
\item {6.} T. Kimura, Prog. Theor. Phys. 81 (1989) 1109; N.
Nakanishi, Int. J. Mod. Phys. A4 (1989) 1055; N. Imai, K. Ishikawa
and I. Tanaka, Prog. Theor. Phys. 81 (1989) 758.
\item {7.} J.M.F. Labastida and A.V. Ramallo, Phys. Lett. B227 (1989)
92.
\item {8.} P.A.M. Dirac, Can. J. Math. 2 (1950) 129; {\it Lectures
on quantum mechanics} (Yeshiva University, New York, 1964).
\item {9.} Q. Lin and G. Ni, Class. Quantum Grav. 7 (1990) 1261.
\item {10.} J.M. Mart\\nez-Fern\'andez and C. Wotzasek, Z. Phys. C43
(1989) 305.
\item {11.} L.D. Faddeev and R. Jackiw, Phys. Rev. Lett. 60 (1988)
1692; J. Govaerts, Int. J. Mod. Phys. A5 (1990) 3625
\item {12.} For a general review on this subject see, for example,
N. Woodhouse, {\it Geometric quantization} (Clarendon Press, Oxford,
1980).
\item {13.}  The treatment of the  symplectic formalism with
constraints we are using was developed in J. Barcelos-Neto and C.
Wotzasek, Mod.  Phys. Lett. A7 (1992) 1737; Int. J. Mod. Phys. A7
(1992) 4981; ; H.  Montani, {\it Sympletic analysis of constrained
systems}, Preprint UFRJ/93.
\item {14.} Other examples of the  symplectic formalism applied in
systems with constraints can be found in M.M. Horta-Barreira and C.
Wotzasek, Phys. Rev. D45 (1992) 1410; J. Barcelos-Neto and E.S.
Cheb-Terrab, Z. Phys. C45 (1992) 133; J. Barcelos-Neto, {\it
Constraints and the hidden symmetry in $2D$ gravity}, to appear in
Phys. Rev. D.
\item {15.} R. Floreanini and R. Jackiw, Phys. Rev. Lett. 59 (1987)
1873.

\bye